\documentclass{article}

\usepackage{arxiv}

\usepackage[utf8]{inputenc} 
\usepackage[T1]{fontenc}    
\usepackage{libertinus}     
\usepackage[scaled=0.85]{inconsolata} 
\usepackage{hyperref}       
\usepackage{url}            
\usepackage{booktabs}       
\usepackage{amsfonts}       
\usepackage{nicefrac}       
\usepackage{microtype}      
\usepackage{graphicx}
\usepackage{subcaption}
\usepackage[numbers,sort&compress]{natbib}
\usepackage{doi}
\usepackage{siunitx}
\usepackage{xcolor}
\usepackage{listings}
\usepackage{tcolorbox}
\tcbuselibrary{listings,skins}

\definecolor{codebg}{HTML}{F5F5F5}
\definecolor{codeframe}{HTML}{DCDCDC}
\definecolor{tableheader}{HTML}{3B5998}
\definecolor{tableheadertext}{HTML}{FFFFFF}
\definecolor{tablerowalt}{HTML}{F2F5FA}
\definecolor{tableborder}{HTML}{B0BEC5}
\definecolor{codegreen}{HTML}{2E7D32}
\definecolor{codegray}{HTML}{6A6A6A}
\definecolor{codeblue}{HTML}{1565C0}
\definecolor{codepurple}{HTML}{7B1FA2}

\usepackage{colortbl}
\newcommand{\thc}[1]{\textcolor{tableheadertext}{\bfseries #1}}

\lstdefinestyle{pythonstyle}{
  language=Python,
  basicstyle=\small\ttfamily,
  keywordstyle=\color{codeblue},
  stringstyle=\color{codegreen},
  commentstyle=\color{codegray}\itshape,
  showstringspaces=false,
  breaklines=true,
  columns=flexible,
}
\lstdefinestyle{shellstyle}{
  basicstyle=\small\ttfamily,
  showstringspaces=false,
  breaklines=true,
  columns=flexible,
}

\newtcblisting{pyblock}{
  listing engine=listings,
  listing options={style=pythonstyle},
  colback=codebg,
  colframe=codeframe,
  left=3mm, right=3mm, top=2mm, bottom=2mm,
  arc=1.5mm,
  boxrule=0.4pt,
  listing only,
}
\newtcblisting{shellblock}{
  listing engine=listings,
  listing options={style=shellstyle},
  colback=codebg,
  colframe=codeframe,
  left=3mm, right=3mm, top=2mm, bottom=2mm,
  arc=1.5mm,
  boxrule=0.4pt,
  listing only,
}

\title{IT-DPC-SRI: A Cloud-Optimized Archive of Italian Radar Precipitation (2010--2025)}

\author{
	Gabriele Franch\thanks{Corresponding author.} \\
	Data Science for Industry and Physics\\
	Fondazione Bruno Kessler, Trento, Italy \\
	\texttt{franch@fbk.eu}
\And
	Elena Tomasi \\
	Data Science for Industry and Physics\\
	Fondazione Bruno Kessler, Trento, Italy
\And
	Uladzislau Azhel \\
	Data Science for Industry and Physics\\
	Fondazione Bruno Kessler, Trento, Italy
\And
	Giacomo Tomezzoli \\
	Data Science for Industry and Physics\\
	Fondazione Bruno Kessler, Trento, Italy
\And
	Alessandro Camilletti \\
	Data Science for Industry and Physics\\
	Fondazione Bruno Kessler, Trento, Italy
\And
	Virginia Poli \\
	ARPAE Emilia-Romagna\\
	Agenzia Italiameteo, Bologna, Italy
\And
	Renata Pelosini \\
	Agenzia Italiameteo\\
	Bologna, Italy
\And
	Gianfranco Vulpiani \\
	Dipartimento di Protezione Civile\\
	Roma, Italy
\And
	Gabriella Scipione \\
	CINECA\\
	Bologna, Italy
\And
	Giuseppe Trotta \\
	CINECA\\
	Bologna, Italy
\And
	Matteo Angelinelli \\
	CINECA\\
	Bologna, Italy
\And
	Leif Denby \\
	Danish Meteorological Institute\\
	Copenhagen, Denmark
\And
	Irene Livia Kruse \\
	Danish Meteorological Institute\\
	Copenhagen, Denmark
\And
	Marco Cristoforetti \\
	Data Science for Industry and Physics\\
	Fondazione Bruno Kessler, Trento, Italy
}

\renewcommand{\shorttitle}{Italian Radar Precipitation Archive}

\hypersetup{
pdftitle={IT-DPC-SRI: A Cloud-Optimized Archive of Italian Radar Precipitation (2010--2025)},
pdfsubject={Weather Radar, Precipitation, Open Data, ARCO, Zarr},
pdfauthor={Gabriele Franch},
pdfkeywords={weather radar, precipitation, Italy, ARCO, Zarr, open data, machine learning},
}

\begin{document}
\maketitle

\begin{abstract}
We present IT-DPC-SRI, the first publicly available long-term archive of Italian weather radar precipitation estimates, spanning 16 years (2010--2025). The dataset contains Surface Rainfall Intensity (SRI) observations from the Italian Civil Protection Department's national radar mosaic, harmonized into a coherent Analysis-Ready Cloud-Optimized (ARCO) Zarr datacube. The archive comprises over one million timesteps at temporal resolutions from 15 to 5 minutes, covering a $1200\times1400$\,\si{\kilo\meter} domain at \SI{1}{\kilo\meter} spatial resolution, compressed from \SI{7}{\tera\byte} to \SI{51}{\giga\byte} on disk. We address the historical fragmentation of Italian radar data---previously scattered across heterogeneous formats (OPERA BUFR, HDF5, GeoTIFF) with varying spatial domains and projections---by reprocessing the entire record into a unified store. The dataset is accessible as a static versioned snapshot on Zenodo, via cloud-native access on the ECMWF European Weather Cloud, and as a continuously updated live version on the ArcoDataHub platform. This release fills a significant gap in European radar data availability, as Italy does not participate in the EUMETNET OPERA pan-European radar composite. The dataset is released under a CC BY-SA 4.0 license.
\end{abstract}


\section{Introduction}
\label{sec:intro}

Weather radar networks provide critical observations for monitoring precipitation at high spatial and temporal resolution. These observations support a wide range of applications, from short-term precipitation forecasting (nowcasting) and flood warning systems to climate research and, increasingly, the training of data-driven weather prediction models. While radar data archives are becoming more accessible in several countries and through pan-European initiatives, significant gaps remain in the availability of long-term, analysis-ready radar datasets.

Italy operates one of the densest national weather radar networks in Europe, covering a territory characterized by a wide variety of topographic areas ranging from lowlands (e.g., the Po Valley), mountainous terrain (with the Alps as the northern frontier and the Apennines as the backbone of the peninsula), and hills. The network is managed by the Dipartimento della Protezione Civile (DPC, Italian Civil Protection Department) through its Centro Funzionale Centrale (CFC) and currently comprises 26 radar systems---20 C-band and 6 X-band---managed by 13 administrations including the DPC, regional authorities, ENAV (Italian air navigation service provider), and CNMCA (National Centre for Aeronautical Meteorology and Climatology). Product generation is centralized at the national level by the DPC, whose operational processing chain produces several projected gridded products, including the Surface Rainfall Intensity (SRI)---an estimate of ground-level precipitation intensity across the Italian territory.

Despite the operational maturity of the Italian radar network, access to historical radar data has been challenging for the research community. A previous effort released TAASRAD19~\citep{franch2020taasrad19}, a high-resolution reflectivity dataset from a single C-band radar in the Trentino--Alto Adige region, demonstrating the value of open radar archives for nowcasting research. The present work extends this approach to the entire national composite, providing a unified precipitation archive covering all of Italy. The DPC has provided real-time access to radar products through open web services since approximately 2020~\citep{dpc2019}, but obtaining historical archives required individual data requests and involved working with data in multiple formats that evolved over time. Early archives were stored in OPERA BUFR format until mid-2020, transitioning to HDF5 thereafter, while the web API serves data as GeoTIFF files. Furthermore, the spatial domain and map projection of the products changed during the 16-year period, making it difficult to construct a consistent time series without significant preprocessing effort.

A further complication arises from Italy's position in the European radar data landscape. The EUMETNET OPERA programme~\citep{saltikoff2019} coordinates the exchange of weather radar data among European national meteorological services and produces pan-European radar composites combining data from over 160 radars. However, Italy is not a member of the OPERA consortium, meaning that the OPERA composites have a gap over Italian territory. This limits the utility of OPERA products for applications requiring coverage of the Italian peninsula and surrounding maritime areas, and motivates the need for dedicated Italian radar data resources. The dataset presented here directly addresses this gap by providing Italian radar data in a format that is technically compatible with OPERA products, enabling users to combine them for complete coverage of continental Europe.

In this paper, we present a new dataset that addresses these challenges: a 16-year archive (2010--2025) of Italian radar SRI observations, harmonized into a single Analysis-Ready Cloud-Optimized (ARCO) datacube~\citep{carver2023}. The dataset is structured as a Zarr store~\citep{zarr2025}, enabling efficient partial reads over the network without downloading the entire archive. We provide both a continuously updated live version through the ArcoDataHub platform and a static snapshot via mlcast-dataset python package and on Zenodo for reproducibility. To our knowledge, this is the first publicly available long-term Italian radar precipitation archive in an analysis-ready format.

\section{Data Source and Original Products}
\label{sec:source}

\subsection{The Italian National Radar Network}

The Italian national weather radar network is coordinated by the Dipartimento della Protezione Civile under the Presidenza del Consiglio dei Ministri. The network data is processed at the Centro Funzionale Centrale (CFC) using operational chains developed in collaboration with the national and international research community, with IT support from Leonardo S.p.A. The processing system, known as WIDE (Weather Ingestion Data Engine), ingests raw polar volume data from individual radars and produces several derived products.

The network currently comprises 26 radar systems---20 C-band and 6 X-band---managed by 13 administrations. The DPC directly operates 7 C-band and 4 X-band systems, all with dual-polarization capability; the remaining radars are operated by regional authorities, ENAV, and CNMCA. C-band is the standard frequency band for operational weather surveillance in Europe, while the X-band systems provide higher resolution coverage in specific areas. Not all radars were operational throughout the entire 16-year period covered by this archive; the network has expanded significantly since 2010.

\subsection{The Surface Rainfall Intensity Product}

The national mosaic generates in real time several projected gridded products: the constant altitude plan position indicator (CAPPI) at altitudes from 2 to \SI{8}{\kilo\meter} spaced by \SI{1}{\kilo\meter}, the vertical maximum indicator (VMI), the Surface Rainfall Intensity (SRI), and its gauge-adjusted version (SRIadj). The SRI product is a gridded estimate of precipitation intensity at the ground surface, expressed in millimeters per hour (mm/h). Unlike simpler radar reflectivity products (such as VMI, which shows the maximum reflectivity in each vertical column), the SRI incorporates additional processing steps to provide a more accurate estimate of actual rainfall reaching the ground.

The dataset presented in this article uses the SRI product, not the gauge-adjusted SRIadj. While SRIadj further refines precipitation estimates by combining radar data with the national rain gauge network, the SRI product relies on radar observations and their processing chain alone, which includes corrections for known radar artifacts such as beam overshooting, bright band effects, and signal attenuation.

In the current operational configuration, the SRI product is updated every 5 minutes. Historical data have varying temporal resolution: prior to 2020, updates occurred at 10-minute to 15-20-minute intervals depending on the period. The spatial resolution has been \SI{1}{\kilo\meter} in the current configuration, though earlier periods used different grid specifications.

\subsection{Historical Data Formats and Challenges}

Accessing historical Italian radar data has been complicated by the evolution of data formats and specifications over the 16-year period. Three main eras can be identified:

The first period (approximately 2010--2020) saw data archived primarily in OPERA BUFR format, following the conventions used by the EUMETNET OPERA programme for radar data exchange. While standardized, BUFR is a binary format that requires specialized software libraries to decode and is not well-suited for analysis workflows based on array-oriented scientific computing.

The second period (2020 onwards) transitioned to HDF5 format following the OPERA Data Information Model (ODIM) conventions. HDF5 is more accessible than BUFR but still requires specific reading code and does not support efficient partial reads over the network.

The operational web API, available since approximately 2020, serves data as GeoTIFF files. This format is widely supported but individual file downloads do not scale well for accessing long time series, and the API provides access primarily to recent data rather than deep archives.

Beyond format heterogeneity, the spatial specifications also changed. The geographic extent, grid spacing, and map projection of the SRI product evolved as the network and processing systems were upgraded. This means that naive concatenation of historical files would produce an inconsistent dataset with spatial misalignments.

\section{Dataset Description}
\label{sec:dataset}

\subsection{Spatial and Temporal Coverage}

The dataset covers the period from January 1, 2010 to December 31, 2025, containing 1,039,785 timesteps. The temporal resolution changed twice during this period: 15-minute intervals from 2010 to June 2014, 10-minute intervals from June 2014 to June 2020, and 5-minute intervals from July 2020 onward. Not all timesteps contain valid data due to occasional system outages, maintenance periods, and the varying operational schedules of the network during the early years; overall completeness is 98.5\%.

The spatial domain covers the Italian peninsula and surrounding areas, extending approximately from \SI{35}{\degree}N to \SI{48}{\degree}N in latitude and from \SI{5}{\degree}E to \SI{20}{\degree}E in longitude. The grid dimensions are 1200 cells in the x-direction and 1400 cells in the y-direction, with a cell size of \SI{1}{\kilo\meter} in both directions.

\subsection{Coordinate Reference System}

All data have been reprojected to a common Transverse Mercator coordinate reference system with the following parameters:

\begin{table}[h]
\caption{Coordinate Reference System Parameters}
\centering
\rowcolors{2}{tablerowalt}{white}
\begin{tabular}{ll}
\arrayrulecolor{tableborder}\toprule
\rowcolor{tableheader} \thc{Parameter} & \thc{Value} \\
\midrule
Projection & Transverse Mercator \\
Latitude of origin & \SI{42}{\degree}N \\
Central meridian & \SI{12.5}{\degree}E \\
Scale factor & 1.0 \\
False easting & \SI{0}{\meter} \\
False northing & \SI{0}{\meter} \\
Ellipsoid & WGS 84 \\
\bottomrule
\end{tabular}
\label{tab:crs}
\end{table}

This projection is centered on the Italian peninsula and minimizes distortion over the domain of interest. The proj4 string for this CRS is:
\begin{shellblock}
+proj=tmerc +lat_0=42 +lon_0=12.5 +k=1 +x_0=0 +y_0=0 +ellps=WGS84 +units=m +no_defs
\end{shellblock}

The dataset includes both the projected coordinates (x, y in meters) as dimension coordinates and the geographic coordinates (latitude, longitude in degrees) as auxiliary 2D coordinate variables.

\subsection{Variables and Units}

The primary data variable is the precipitation rate, named \texttt{RR}. The values are stored in units of \si{\milli\meter\per\hour} (equivalent to \si{\kilo\gram\per\meter\squared\per\hour}), consistent with the standard meteorological convention for precipitation rate products.

Missing data are represented as NaN (Not a Number) values, which are the standard convention for floating-point missing data in xarray-based workflows~\citep{hoyer2017}. The data are stored as 32-bit floating-point values.

\subsection{File Format and Structure}

The dataset is stored in Zarr format~\citep{zarr2025}, a chunked, compressed, N-dimensional array format designed for cloud-native workflows. Each chunk is compressed using the Blosc~\citep{alted2010} meta-compressor with the Zstd codec at level~9, achieving a 136:1 compression ratio (from \SI{7.0}{\tera\byte} uncompressed to \SI{51}{\giga\byte} on disk). Zarr supports efficient partial reads, meaning that users can load only the spatial or temporal subset they need without downloading the entire archive.

\begin{table}[h]
\caption{Zarr Store Structure}
\centering
\rowcolors{2}{tablerowalt}{white}
\begin{tabular}{lll}
\arrayrulecolor{tableborder}\toprule
\rowcolor{tableheader} \thc{Variable} & \thc{Dimensions} & \thc{Chunks} \\
\midrule
\texttt{RR} & (time, y, x) & (1, 1400, 1200) \\
\texttt{time} & (time,) & (1039785,) \\
\texttt{x} & (x,) & (1200,) \\
\texttt{y} & (y,) & (1400,) \\
\texttt{lat} & (y, x) & (1400, 1200) \\
\texttt{lon} & (y, x) & (1400, 1200) \\
\texttt{missing\_times} & (missing\_times,) & (15530,) \\
\texttt{crs} & () & () \\
\bottomrule
\end{tabular}
\label{tab:zarr}
\end{table}

The chunking strategy stores each timestep as a single chunk in space (the full 1400$\times$1200 grid), which is optimized for spatial operations on individual timesteps---the most common access pattern for nowcasting and similar applications. A dedicated \texttt{missing\_times} coordinate lists the 15,530 timestamps for which no valid observation is available, allowing users to quickly identify data gaps without scanning the full data array.

\subsection{Variable Attributes and Grid Mapping}

The precipitation variable \texttt{RR} carries CF-compliant metadata~\citep{cfconventions2024} linking it to the coordinate reference system and describing its physical quantity. The scalar variable \texttt{crs} serves as a \emph{grid mapping} variable following the CF conventions: it holds no data values itself, but its attributes encode the full definition of the projected coordinate reference system. This allows CF-aware tools (including xarray and GDAL) to automatically interpret the spatial coordinates and perform reprojection or overlay with other geospatial datasets.

Table~\ref{tab:attrs} lists selected attributes of these two variables.

\begin{table}[h]
\caption{Key Variable Attributes}
\centering
\rowcolors{2}{tablerowalt}{white}
\begin{tabular}{lll}
\arrayrulecolor{tableborder}\toprule
\rowcolor{tableheader} \thc{Variable} & \thc{Attribute} & \thc{Value} \\
\midrule
\texttt{RR} & \texttt{long\_name} & Total precipitation rate \\
\texttt{RR} & \texttt{standard\_name} & \texttt{rainfall\_flux} \\
\texttt{RR} & \texttt{units} & kg\,m$^{-2}$\,h$^{-1}$ \\
\texttt{RR} & \texttt{grid\_mapping} & \texttt{crs} \\
\texttt{RR} & \texttt{\_FillValue} & NaN \\
\texttt{RR} & compressor & Blosc (Zstd, level 9) \\
\midrule
\texttt{crs} & \texttt{grid\_mapping\_name} & \texttt{transverse\_mercator} \\
\texttt{crs} & \texttt{latitude\_of\_projection\_origin} & 42.0 \\
\texttt{crs} & \texttt{longitude\_of\_central\_meridian} & 12.5 \\
\texttt{crs} & \texttt{scale\_factor\_at\_central\_meridian} & 1.0 \\
\texttt{crs} & \texttt{reference\_ellipsoid\_name} & WGS 84 \\
\texttt{crs} & \texttt{GeoTransform} & $-600000,\;1000,\;0,\;650000,\;0,\;-1000$ \\
\bottomrule
\end{tabular}
\label{tab:attrs}
\end{table}

The \texttt{grid\_mapping} attribute on \texttt{RR} points to the \texttt{crs} variable, establishing the link between the data array and its spatial reference. The \texttt{GeoTransform} attribute encodes the affine transformation from pixel coordinates to projected coordinates (origin at $x=-600\,000$\,m, $y=650\,000$\,m; pixel size \SI{1000}{\meter}), following the GDAL convention.

\section{Data Processing and Harmonization}
\label{sec:processing}

\subsection{Source Data Collection}

Historical radar data were obtained through a concerted retrieval effort between Dipartimento della Protezione Civile, ARPAE Emilia-Romagna and Agenzia Italiameteo. The archives were provided in their original formats: OPERA BUFR for the earlier period, HDF5/ODIM for the later period. Real-time data since 2024 were collected using DPC FTP distribution service that allows to retrieve the same products published in the official public REST API and WebSocket notification services provided by the Radar-DPC platform~\citep{dpc2019}.

\subsection{Spatial Harmonization}

A key challenge was harmonizing data from different periods that used different spatial specifications. We adopted the current operational grid (1200$\times$1400 at \SI{1}{\kilo\meter} resolution in the Transverse Mercator projection described in Section~\ref{sec:dataset}) as the target grid and reprojected all historical data to this common specification.

Reprojection was performed using the least possible destructive transformations for data values and appropriate handling of the precipitation rate units. Quality flags and other auxiliary information from the source products (when available) were not preserved in the harmonized archive to maintain a simple, streamlined data structure focused on the precipitation estimates.

\subsection{Temporal Alignment}

The varying temporal resolution of historical data required alignment to a common time grid. We adopted the current 5-minute interval as the target, with timestamps at exact multiples of 5 minutes UTC (00, 05, 10, \ldots, 55 minutes past each hour).

For periods with coarser temporal resolution (15-minute intervals before June 2014; 10-minute intervals from June 2014 to June 2020), the available observations were assigned to the nearest target timestep, with the remaining timesteps marked as missing (NaN). This approach preserves the original sampling without artificial interpolation of precipitation rates, which would be physically inappropriate for such a highly variable quantity.

\subsection{Quality Considerations}

The harmonized dataset inherits the quality characteristics of the source SRI product, including its strengths (multi-sensor fusion) and limitations (residual radar artifacts, coverage gaps in mountainous regions, varying network density over time).

We performed minimal quality filtering beyond the processing already applied in the operational chain. Users should be aware that data quality and coverage improve substantially from 2010 to the present as the network expanded and processing algorithms were refined.

\section{Dataset Characterization}
\label{sec:characterization}

This section presents a quantitative characterization of the harmonized archive, including the spatial domain, temporal coverage, data completeness, precipitation statistics, and seasonal variability.

\subsection{Spatial Domain}

\begin{figure}[htbp]
\centering
\includegraphics[width=13cm]{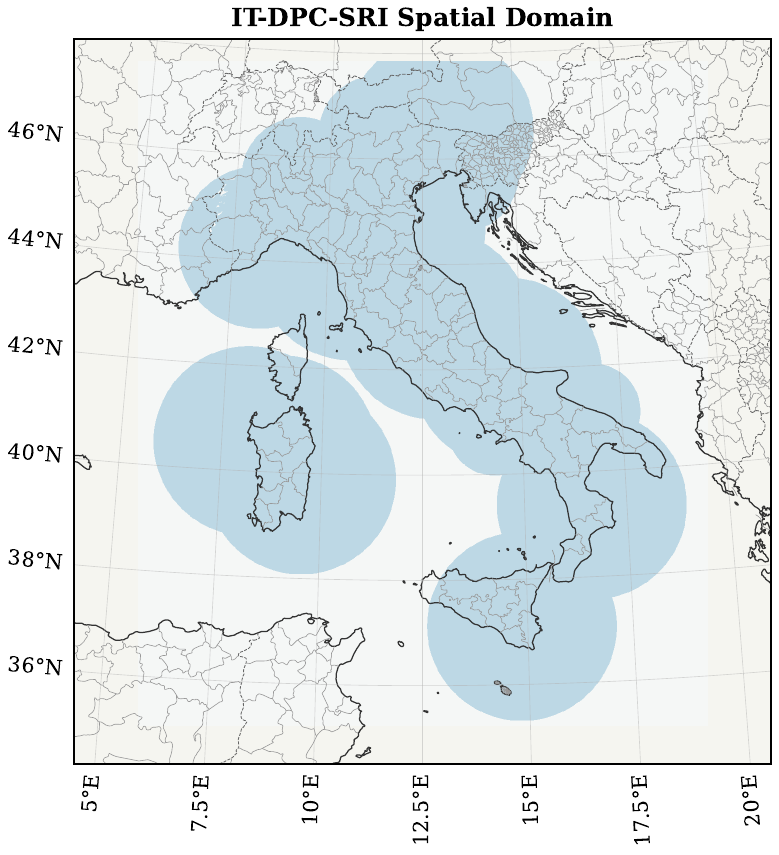}
\caption{Spatial domain of the IT-DPC-SRI dataset. The shaded region shows the radar coverage footprint (grid cells with at least one valid observation). Country borders and coastlines from Natural Earth (10\,m resolution).}
\label{fig:domain}
\end{figure}

Figure~\ref{fig:domain} shows the spatial extent of the dataset. The radar footprint---defined as the set of grid cells that contain at least one valid observation---covers the Italian peninsula, the major islands (Sicily and Sardinia), and portions of the surrounding seas and neighboring countries. The domain is projected onto a Transverse Mercator grid centered at \SI{42}{\degree}N, \SI{12.5}{\degree}E (see Table~\ref{tab:crs}), with dimensions of 1400 cells (north--south) by 1200 cells (east--west) at \SI{1}{\kilo\meter} resolution.

\subsection{Temporal Coverage}

The archive spans from January 1, 2010 to December 31, 2025, containing 1,039,785 timesteps with 15,530 missing (1.5\%). The temporal resolution changed twice during this period: 15-minute intervals from 2010 to June 2014, 10-minute intervals from June 2014 to June 2020, and 5-minute intervals from July 2020 onward.

\begin{figure}[htbp]
\centering
\includegraphics[width=\columnwidth]{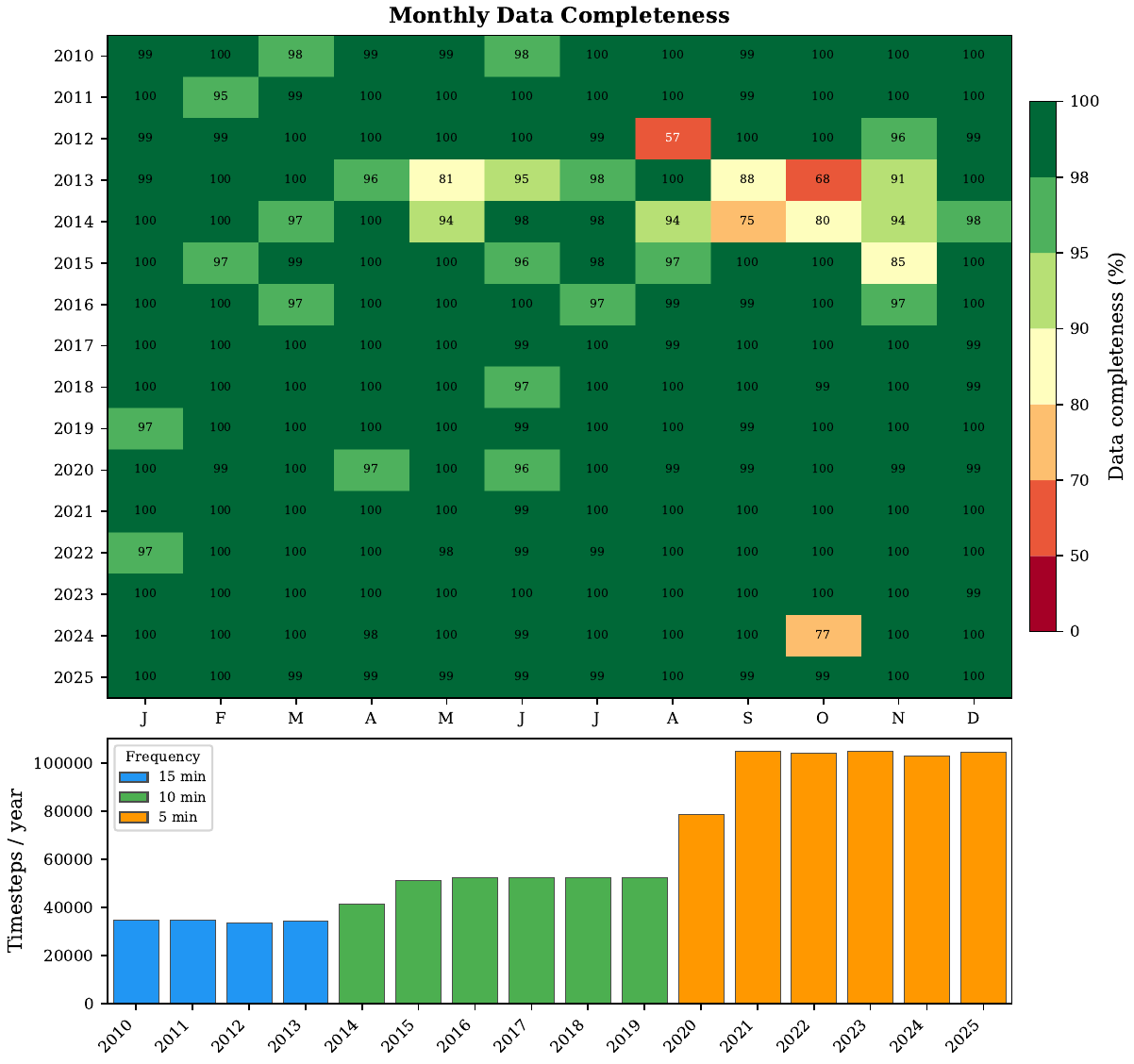}
\caption{Temporal coverage of the dataset. \textbf{Top:} year-by-month completeness heatmap (fraction of expected timesteps present). \textbf{Bottom:} total timesteps per year, colored by temporal frequency era (15\,min in blue, 10\,min in orange, 5\,min in green).}
\label{fig:temporal_coverage}
\end{figure}

Figure~\ref{fig:temporal_coverage} presents two views of temporal completeness. The upper panel shows a year-by-month heatmap of data availability, expressed as the fraction of expected timesteps that are actually present. The lower panel shows the total number of timesteps per year, colored by the temporal frequency era. The transition to higher temporal frequency in 2020 is clearly visible as a sharp increase in the annual timestep count.

\subsection{Spatial Coverage}

\begin{figure}[htbp]
\centering
\includegraphics[width=12cm]{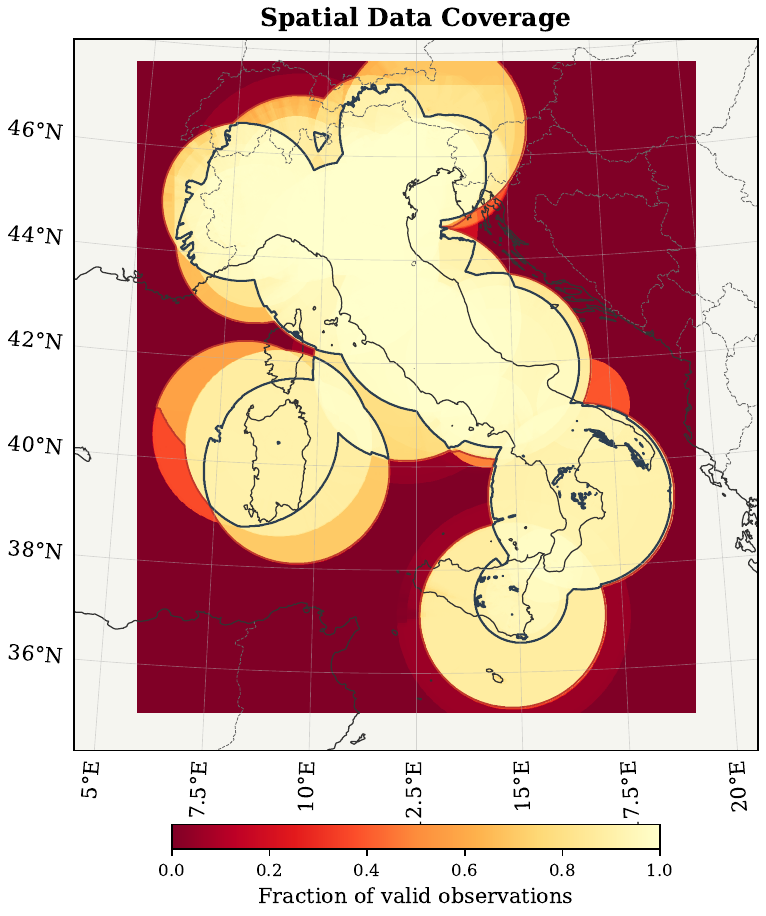}
\caption{Fraction of valid (non-NaN) observations per grid cell, computed from 1000 uniformly spaced timesteps. Contour lines are drawn at 0.5 and 0.9 fractional coverage.}
\label{fig:spatial_coverage}
\end{figure}

Not all grid cells have valid data at every timestep. The spatial coverage varies due to radar range limitations, terrain shadowing in mountainous regions, and network expansion over time. Figure~\ref{fig:spatial_coverage} maps the fraction of sampled timesteps for which each grid cell contains a valid (non-NaN) observation, computed from 1000 uniformly spaced timesteps across the archive.

Coverage is highest (${>}90\%$) over the Po Valley and central Italy, where multiple overlapping radar beams provide robust detection. Coverage degrades in peripheral areas and over the sea at distances exceeding the typical ${\sim}$\SI{150}{\kilo\meter} radar range. Some Alpine valleys and part of Sicily show reduced coverage due to beam blockage by complex terrain and missing radar coverage.

\subsection{Example Precipitation Event}

\begin{figure*}[htbp]
\centering
\includegraphics[width=\textwidth]{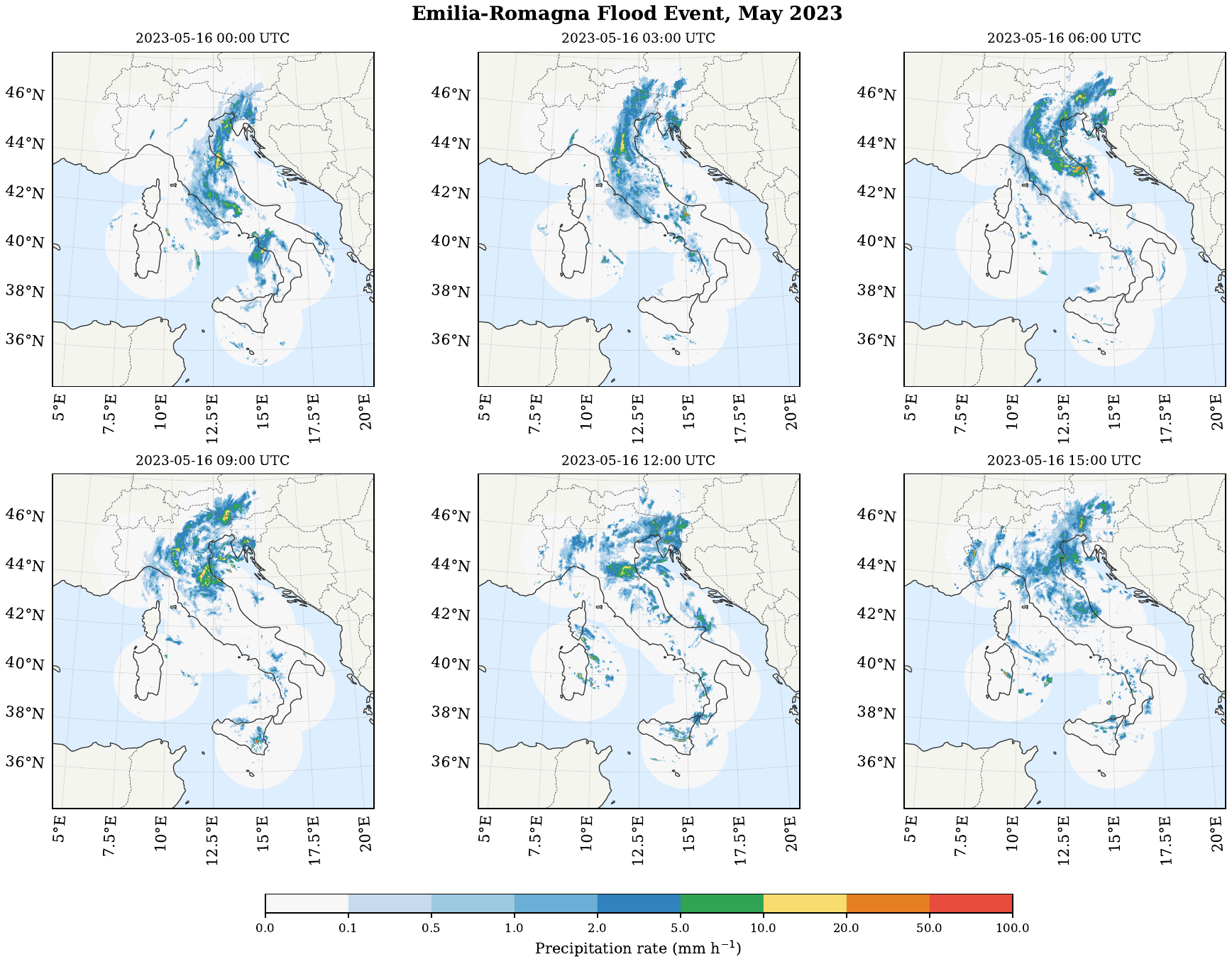}
\caption{Six consecutive precipitation frames from the Emilia-Romagna flood event (May 16--17, 2023). The panels show precipitation rate in mm\,h$^{-1}$ at 3-hour intervals, illustrating the evolution and fine spatial structure of intense rainfall.}
\label{fig:precipitation_event}
\end{figure*}

To illustrate the spatial detail captured by the dataset, Figure~\ref{fig:precipitation_event} shows six consecutive frames from the Emilia-Romagna flood of May 16--17, 2023. This event caused catastrophic flooding in north-central Italy and was one of the most significant hydrometeorological events in the archive. The panels show the evolution of an organized precipitation system as it moved across the region, with rainfall rates exceeding \SI{50}{\milli\meter\per\hour} in the most intense cells.

\subsection{Precipitation Statistics}

\begin{figure*}[htbp]
\centering
\includegraphics[width=\textwidth]{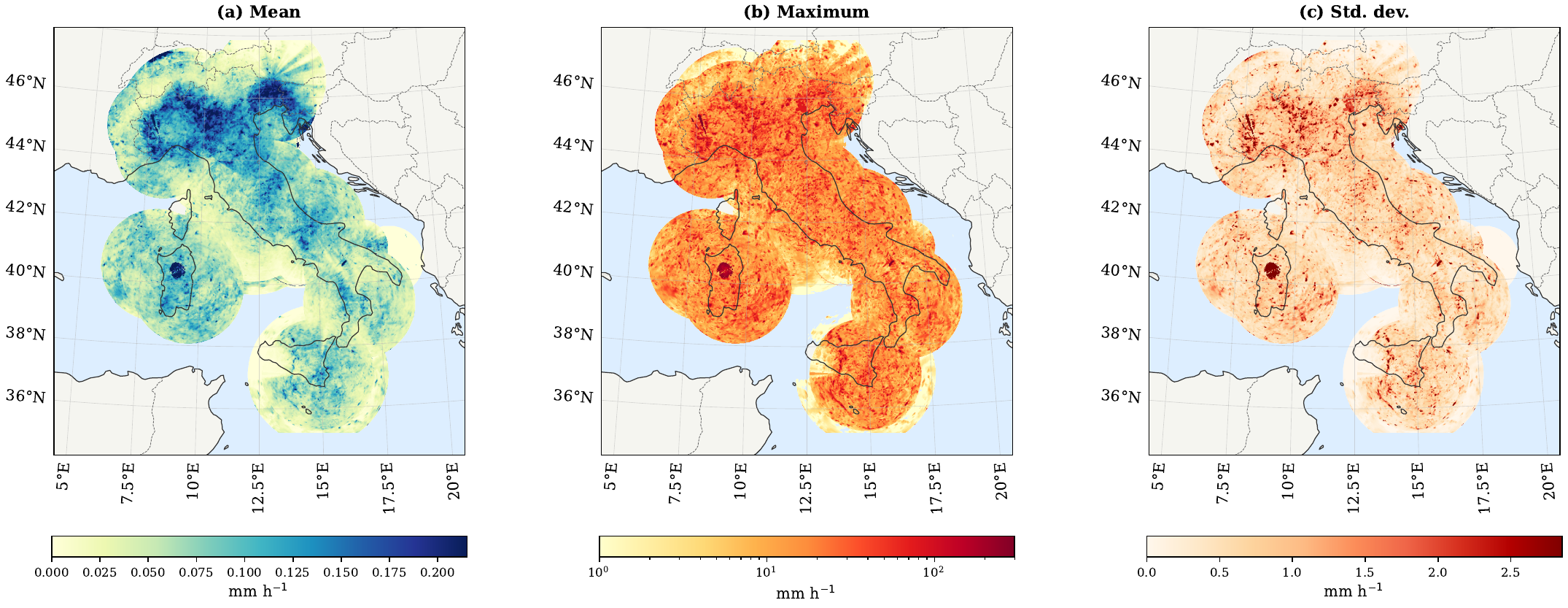}
\caption{Spatial maps of precipitation statistics computed from 2000 sampled timesteps. \textbf{(a)}~Temporal mean, \textbf{(b)}~maximum observed rate, \textbf{(c)}~standard deviation. All values in mm\,h$^{-1}$.}
\label{fig:precip_stats}
\end{figure*}

Figure~\ref{fig:precip_stats} presents spatial maps of three precipitation statistics computed from 2000 uniformly sampled timesteps using Welford's online algorithm: the temporal mean precipitation rate, the maximum observed rate, and the standard deviation.

\begin{figure}[htbp]
\centering
\includegraphics[width=13 cm]{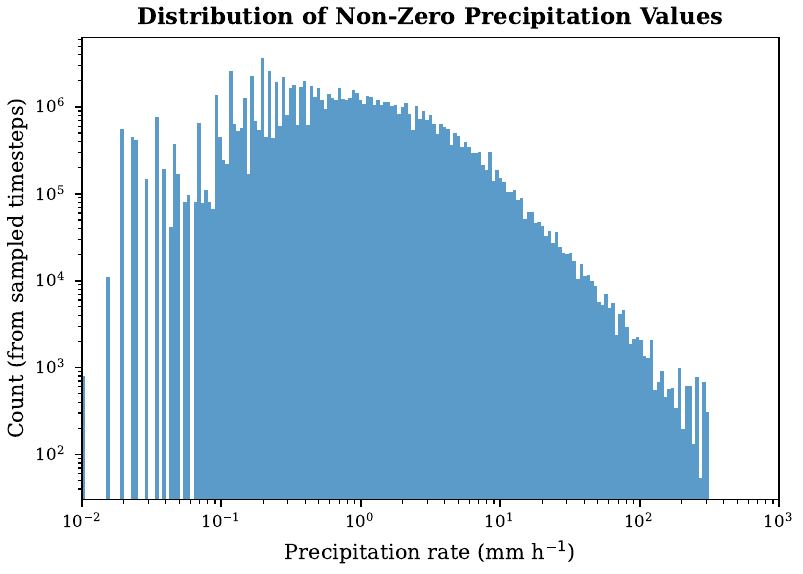}
\caption{Log--log probability density of non-zero precipitation values (mm\,h$^{-1}$), computed from 2000 sampled timesteps.}
\label{fig:precip_histogram}
\end{figure}

Figure~\ref{fig:precip_histogram} shows the probability density of non-zero precipitation values on a log--log scale. The distribution spans several orders of magnitude, from drizzle at ${\sim}0.01$\,mm\,h$^{-1}$ to extreme rates exceeding \SI{100}{\milli\meter\per\hour}. The approximate power-law behavior in the tail is consistent with the known heavy-tailed character of precipitation intensity distributions.

\subsection{Seasonal Cycle}

\begin{figure}[htbp]
\centering
\includegraphics[width=14cm]{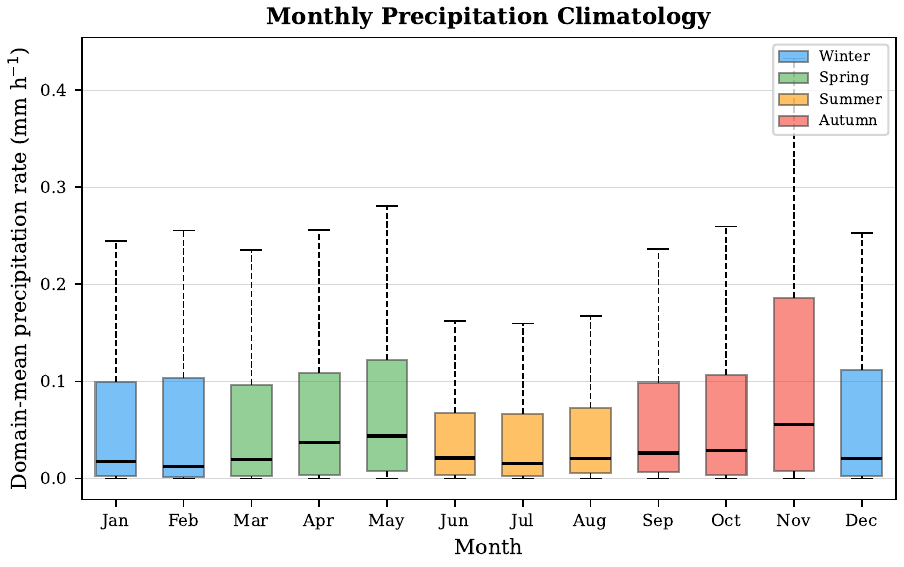}
\caption{Monthly distribution of domain-mean precipitation rate (mm\,h$^{-1}$), based on 3000 sampled timesteps. Boxes show the interquartile range with median line; whiskers extend to 1.5$\times$IQR. Colors indicate meteorological seasons: winter (blue), spring (green), summer (orange), autumn (red).}
\label{fig:monthly_cycle}
\end{figure}

Figure~\ref{fig:monthly_cycle} shows the monthly distribution of domain-mean precipitation rate, based on 3000 uniformly sampled timesteps. The box plot reveals a clear seasonal cycle with higher precipitation rates during late spring and autumn (May and November) and lower rates during winter and especially summer. This pattern is consistent with the known Mediterranean precipitation regime of Italy, where convective activity peaks in late spring and cyclonic systems bring widespread rainfall in autumn.

\subsection{Dataset Summary}

Table~\ref{tab:summary} provides a compact overview of the key dataset properties.

\begin{table}[h]
\caption{Summary of the IT-DPC-SRI dataset.}
\centering
\rowcolors{2}{tablerowalt}{white}
\begin{tabular}{ll}
\arrayrulecolor{tableborder}\toprule
\rowcolor{tableheader} \thc{Property} & \thc{Value} \\
\midrule
Time range & 2010-01-01 to 2025-12-31 \\
Total timesteps & 1,039,785 \\
Missing timesteps & 15,530 (1.5\%) \\
Grid dimensions & 1400 $\times$ 1200 pixels \\
Spatial resolution & 1\,km \\
Domain extent & $\sim$1200 $\times$ 1400\,km \\
Data variable & \texttt{RR} (precipitation rate) \\
Units & mm\,h$^{-1}$ (= kg\,m$^{-2}$\,h$^{-1}$) \\
Data type & float32 \\
Chunk layout & (1, 1400, 1200) \\
Compression & Blosc/Zstd, level 9 \\
Uncompressed volume & 7.0\,TB \\
Compressed volume & 51\,GB \\
Compression ratio & 136:1 \\
Temporal frequency & 15min (2010--2014), 10min (2014--2020), 5min (2020--2025) \\
CRS & Transverse Mercator (42$^\circ$N, 12.5$^\circ$E) \\
License & CC--BY--SA--4.0 \\
\bottomrule
\end{tabular}
\label{tab:summary}
\end{table}

\section{Data Access}
\label{sec:access}

The dataset is available through three complementary channels, offering different trade-offs between convenience, reproducibility, and timeliness.

\subsection{Zenodo Download}

For applications requiring strict reproducibility or offline access, the complete 2010--2025 archive is available as a downloadable Zarr store on Zenodo (\doi{10.5281/zenodo.18637608}). Each release is versioned and assigned a DOI, allowing citations to refer to a specific, immutable version of the data. The Zarr store can be opened directly with xarray after decompressing the tar archive:

\begin{pyblock}
import xarray as xr

ds = xr.open_zarr("path/to/italian-radar-dpc-sri.zarr/")
\end{pyblock}

\subsection{Interactive Cloud Access via MLCast Datasets}

The same 2010--2025 archive is hosted on the ECMWF European Weather Cloud S3-compatible object storage, providing interactive cloud-native access without requiring a full download. The data are publicly accessible with anonymous read permissions and no authentication is required.

The simplest way to access the dataset is through the \texttt{mlcast-datasets} Python package, which provides an intake catalog of community-curated datasets for machine learning--based nowcasting, developed within the EUMETNET E-AI Optional Programme:

\begin{pyblock}
import mlcast_datasets

cat = mlcast_datasets.open_catalog()
ds = cat.precipitation.it_dpc_sri_5min.to_dask()
\end{pyblock}

Alternatively, the dataset can be accessed directly with xarray and fsspec:

\begin{pyblock}
import xarray as xr

ds = xr.open_zarr(
    "s3://mlcast-source-datasets/IT-DPC-SRI/"
    "v0.1.0/italian-radar-dpc-sri.zarr/",
    storage_options={
        "anon": True,
        "endpoint_url":
            "https://object-store.os-api.cci2.ecmwf.int",
    },
)
\end{pyblock}

Both methods return a lazily-loaded dask-backed xarray Dataset, enabling efficient partial reads of arbitrary spatial and temporal subsets without downloading the full archive.

\subsection{ArcoDataHub --- Live Updated Version}

A continuously updated version of the dataset is available on the ArcoDataHub platform (\url{https://arcodatahub.com}). Unlike the static 2010--2025 archive described above, this version is updated in near real-time: new SRI observations are appended every 5 minutes as they become available from the DPC operational system. The key differences from the static archive are:

\begin{itemize}
\item The time dimension extends to 2050 (prefilled with NaN values), allowing new timesteps to be appended without modifying the store structure.
\item A \texttt{last\_valid} global attribute indicates the timestamp of the most recently ingested observation. Users should check this attribute to determine the actual extent of valid data.
\item Access requires authentication (username and API key), obtainable through the ArcoDataHub registration process.
\end{itemize}

The following example demonstrates accessing the live dataset and computing a 24-hour precipitation accumulation from the most recent data:

\begin{pyblock}
import xarray as xr
import pandas as pd

user = "your_username"
key = "your_api_key"
dataset_name = "italian-radar-dpc-sri.zarr"
url = f"https://{user}:{key}@api.arcodatahub.com/S3/{dataset_name}"

ds = xr.open_zarr(url)

# Get the last valid timestamp
last_valid = pd.Timestamp(ds.attrs["last_valid"])

# Select the last 24 hours
start = last_valid - pd.Timedelta(days=1)
selection = ds.sel(time=slice(start, last_valid))

# Sum precipitation (convert from rate to accumulation)
# Each timestep is 5 minutes = 1/12 hour
# RR is in mm/h, so multiply by 1/12 to get mm
accumulation = selection.sum(dim="time") / 12

# Export to netCDF on local machine
accumulation.to_netcdf("last_24h_accumulation.nc")
\end{pyblock}

For applications requiring strict reproducibility, the Zenodo or European Weather Cloud versions should be preferred, as the live dataset evolves continuously.

\section{Intended Applications}
\label{sec:applications}

The combination of high spatiotemporal resolution, long historical record, and cloud-optimized access makes this dataset suitable for several research and operational domains. Precipitation nowcasting stands out as a primary use case: deep learning approaches require large training datasets to learn the complex spatiotemporal dynamics of precipitation fields, and 16 years of high-frequency radar observations provide a substantial resource for this purpose. The dataset is currently being used within the MLCast initiative and the Italian AI-Factory IT4LIA project for the development of radar-based nowcasting models. Hydrological modeling is another natural application, particularly for flash flood forecasting in the Italian context, where complex terrain and Mediterranean climate patterns produce rapid hydrological responses; the 16-year record includes several significant flood events across Italian river basins.

The archive also supports climate-oriented analyses of precipitation patterns, extremes, and seasonal variability, though users should account for network evolution and calibration changes over time (see Section~\ref{sec:limitations}). More broadly, the chaotic dynamics of precipitation and its heavy-tailed intensity distribution make this dataset a challenging benchmark for machine learning research on spatiotemporal prediction.

\section{Limitations and Known Issues}
\label{sec:limitations}

Users should be aware of several limitations when working with this dataset.

The early years of the archive (2010--2015) have substantially lower data quality and coverage than recent years. The radar network was smaller, some processing algorithms were less mature, and there were more frequent data gaps. Research requiring consistent quality throughout should consider restricting analysis to more recent years or explicitly accounting for these temporal variations.

The SRI product represents estimated surface rainfall intensity, not directly measured reflectivity. The retrieval involves assumptions about the drop size distribution and other factors that introduce uncertainty. Unlike the gauge-adjusted SRIadj product, the SRI used in this dataset relies solely on radar-derived estimates without rain gauge fusion, which may result in larger biases in regions with limited radar coverage or complex terrain.

Spatial coverage is limited by terrain and radar siting. Complex topography in the Alps and Apennines creates shadowing effects and beam blockage that result in degraded coverage or missing data in some mountain valleys. Maritime coverage extends several kilometers offshore but is limited by Earth curvature and the ${\sim}$\SI{150}{\kilo\meter} typical maximum range of the radars.

The dataset contains only the derived precipitation rate product and does not include reflectivity or other radar moments. Users requiring access to raw radar data would need to work directly with DPC archives.

\section{Data Availability and License}
\label{sec:availability}

The dataset is released under the Creative Commons Attribution-ShareAlike 4.0 International (CC BY-SA 4.0) license and is archived on Zenodo~\citep{franch2025}. Users may freely use, modify, and redistribute the data provided that appropriate credit is given to the Dipartimento della Protezione Civile as the data source and to Fondazione Bruno Kessler for the data processing and harmonization. The three access channels (Zenodo, European Weather Cloud, and ArcoDataHub) are described in Section~\ref{sec:access}.

\section{Conclusions}
\label{sec:conclusions}

We have presented a new publicly available archive of Italian weather radar precipitation estimates, spanning 16 years from 2010 to the present. By harmonizing heterogeneous historical data into a unified ARCO datacube, we provide the research community with easy access to a valuable observational resource that was previously difficult to obtain and use.

The dataset addresses a significant gap in European radar data availability, as Italy does not participate in the OPERA pan-European composite. The combination of long historical record, high spatiotemporal resolution, live updating capability, and cloud-optimized format makes this resource suitable for a wide range of applications from operational nowcasting to climate analysis and machine learning research.

We anticipate that this release will enable new research on precipitation patterns in the Italian region and contribute to advancing data-driven approaches to weather prediction and hydrological forecasting.

\section*{Acknowledgments}

We thank the Dipartimento della Protezione Civile for providing access to the radar data archives and for their continued operation of the national radar network. We acknowledge the contributions of the regional meteorological services and other organizations that operate radars contributing to the national composite.

This work was funded by the European Union through the EuroHPC Joint Undertaking under grant agreement No.~101234224 (IT4LIA --- Italy for Artificial Intelligence), within the HORIZON-JU-EUROHPC-2025-AI-01-IBA-01 programme.

\bibliographystyle{unsrtnat}
\bibliography{references}

\end{document}